# Imaging and Dynamics of Light Atoms

# and Molecules on Graphene

Jannik C. Meyer, C. O. Girit, M. F. Crommie, and A. Zettl

*Department of Physics, University of California at Berkeley, and Materials Sciences Division, Lawrence Berkeley National Laboratory, Berkeley, CA 94720 U.S.A.*

**Observing the individual building blocks of matter is one of the primary goals of microscopy. The invention of the scanning tunneling microscope [1] revolutionized experimental surface science in that atomic-scale features on a solid-state surface could finally be readily imaged. However, scanning tunneling microscopy has limited applicability due to restrictions, for example, in sample conductivity, cleanliness, and data aquisition rate. An older microscopy technique, that of transmission electron microscopy (TEM) [2, 3] has benefited tremendously in recent years from subtle instrumentation advances, and individual heavy (high atomic number) atoms can now be detected by TEM [4 - 7] even when embedded within a semiconductor material [8, 9]. However, detecting an individual low atomic number atom, for example carbon or even hydrogen, is still extremely challenging, if not impossible, via conventional TEM due to the very low contrast of light elements [2, 3, 10 - 12]. Here we demonstrate a means to observe, by conventional transmission electron microscopy, even the smallest atoms and molecules: On a clean single-layer graphene membrane, adsorbates such as atomic hydrogen and carbon can be seen as if they were suspended in free space. We directly image such individual adatoms, along with carbon chains and vacancies, and investigate their dynamics in real time. These techniques open a way to reveal dynamics of more complex chemical reactions or identify the atomic-scale structure of unknown adsorbates. In addition, the study of atomic scale defects in graphene may provide insights for nanoelectronic applications of this interesting material.**

The atomic-scale resolution of TEM comes at the price of requiring that the transmitted electron beam reach the imaging lenses and detector, and therefore TEM works only for ultra thin, electron transparent samples. In high-resolution transmission electron microscopy (HRTEM) and all related techniques such as electron diffraction, scanning transmission electron microscopy (STEM), electron energy loss spectroscopy or elemental mapping, any support film or membrane provides a background signal that is most significant for the smallest objects under investigation. Individual nanoscale particles or molecules usually need to be supported by a continuous membrane, as only tubular or rod-shaped nanoparticles (such as carbon nanotubes) can be suspended across holes in the membrane. Indeed, single-walled carbon nanotubes (SWNTs) have been utilized for low-background TEM studies of encapsulated molecules [13 - 15] or defects in the cylinder-shaped graphene sheets [16, 17]. However, the limited space, harsh filling procedures, and strongly curved shape of the sheet limit the applicability and complicate the analysis.

As we demonstrate below, a graphene membrane provides the ultimate sample support for electron microscopy. With a thickness of only one atom, it is the thinnest possible continuous material. Due to its crystalline nature, a graphene support membrane is either completely invisible or, if the graphene lattice is resolved by a very-high-resolution microscope, its contribution to the imaging signal can be easily subtracted. Graphene is also a good electrical conductor and therefore displays minimal charging effects from the electron beam. Remarkably, we find that a graphene membrane enables single adatom sensitivity even when using a common TEM that does not resolve a graphitic lattice.

In order to observe adsorbates at the single-atom level, the graphene support membrane must be exceptionally clean. In contrast to an earlier graphene membrane preparation method [18], our approach does not rely on electron beam lithography and is simple enough to be reproducible in any basic microscopy laboratory (a detailed description of the sample preparation is given in the supplementary information). In brief, we start with graphene cleaved onto a substrate using an adhesive tape [19 - 21] and transfer selected sheets to commercially available TEM grids. We use electron diffraction analysis as described in [18] to verify the presence of a single layer. Figure 1a shows a low magnification view of a graphene sheet suspended across the 1.3 μm holes of the perforated carbon foil, with a close-up shown in Fig. 1b. More than 50% of the area on these graphene membranes appears exceptionally clean, with no dramatic contrast in high-resolution TEM images (Fig. 1b). As we now demonstrate, however, these "clean" regions contain individual adatoms that are readily observable by TEM. Although individual exposures can reveal useful data, a dramatic improvement in the signal-to-noise ratio is achieved by summing multiple subsequent frames (corrected for sample drift), which effectively increases the exposure time beyond the dynamic range of the TEM CCD detector. Summing as few as 5 frames yields striking visual improvement with atomic-scale features (including individual adatoms) becoming readily apparent, and summing 100 frames reduces the noise to below 0.12% (standard deviation in a relatively featureless region of the graphene membrane).

Fig. 2a shows a TEM image in which an individual carbon atom, attached to the graphene membrane, is identified by an arrow. We recorded eight consecutive, essentially identical images to that of Fig. 2a (each a summation of 20 frames on the CCD), demonstrating that the carbon atom did not adsorb or desorb during the time of exposure. To identify the adatom, image simulations were carried out as described in the supplementary information. The good agreement between the TEM data and image simulation (Fig. 2b) confirms the carbon atom identification. Simulations for individual boron, nitrogen or oxygen adatoms also provide reasonable fits, however, carbon is the dominant component of vacuum contamination and surface adsorbates within our TEM, making these other candidates unlikely.

Closer inspection of Fig. 2a indicates also faint atomic-scale structure distinctly different from carbon adatoms. To highlight these faint features, we show in Fig. 2d a summation of 100 consecutive TEM frames for the same physical region. In this representation the carbon adatoms display sharp contrast at the saturation of the gray level scale. Fig. 2d reveals a moderate density of additional dark features (dark gray points, a selection of which are identified with red arrows) with identical intensity profiles, all with a central dip reduction near 0.6% of the mean bright-field intensity (Fig. 2e). By comparing the TEM image data for these additional features to adatom simulations, we rule out any adatom heavier than helium, as well as a substitution of carbon atoms in the graphene membrane by other elements. However, a hydrogen adatom results in the correct 0.6% dip in the bright-field intensity, shown by the red curve in Fig. 2e. The large number of essentially identical adatom profiles, along with the excellent agreement with the simulated contrast, provide convincing evidence that we have, for the first time, detected individual hydrogen atoms by transmission electron microscopy.

We emphasize that we do not claim to have resolved a hydrogen-carbon distance, which would require advanced aberration-corrected instrumentation. However, detecting an isolated hydrogen atom against a nearly-invisible background only requires an adequate signal-to-noise ratio. Electron scattering from hydrogen has been detected previously in electron diffraction experiments [22] and was found to produce a 3-4 times lower signal than carbon, in agreement with our values. In order to verify the uniqueness of the hydrogen atom identification, we consider whether alternative structures may lead to the observed contrast. The contrast match to simulations, using edges or vacancies in the sheet as reference, is better than a factor of two (see supplementary info) in our experiment. Then, for an adatom on the sheet, only hydrogen or helium can produce the observed contrast. We rule out helium since it will not bind to carbon and is not present anywhere in the experiment. We considered all known defects of graphene sheets. All vacancies, relaxed [23] or not, will produce a white spot. An adatom-vacancy pair [16], even with the minimum separation below the resolution of our microscope, would show a white-and-black symmetric intensity at a detectable level according to simulations and was indeed observed. A Stone-Wales defect [23] is also expected to have a white and black symmetric contrast with zero mean value. Extended defects, such as dislocations, would not produce the rotationally symmetric contrast of an adatom. Finally, a single layer graphene membrane has a much smaller set of possible defects than graphite (or few-layer graphene), e. g. it can not have interstitials or bonds between layers. Thus, we confirm the identification of the hydrogen adatom.

In addition to individual adatoms, we observe by the same TEM imaging methods the generation (by the electron beam) and dynamics of defects (vacancies) in the graphene membrane, as well as the dynamics of a variety of molecular-scale adsorbates. The formation of vacancies due to knock-on damage by the electron beam is shown in Figs. 3a-c. We also observe vacancies that disappear by interaction with mobile adsorbates. Larger adsorbates (small molecules) become trapped preferentially at defects, and can be observed at one position for typically one to five minutes. Frequently, we see that the vacancy disappears along with the trapped adsorbate (Fig. 3d-f), and the missing carbon atom has obviously been resubstituted from the adsorbate. Further, we can directly observe linear molecules on graphene membranes (Fig. 4) that resemble an individual alkane or alkene carbon chain. These molecules are found to spontaneously appear in the field of view, presumably adsorbed onto the graphene membrane from the vacuum contamination. We can follow their dynamics for a few minutes until they decompose in the electron beam, as shown in Fig. 4b-d and in the supplementary videos.

The remarkable TEM imaging capability afforded by a suspended, single graphene membrane warrants further discussion. For an ideal graphene sheet, there are no components in the structure with a period larger than 2.1Å, which is beyond the information limit of approximately 2.9Å for the microscope used in the present studies (JEOL 2010 operated at 100kV). Therefore, although the ideal graphene membrane cannot be resolved under these conditions, any perturbation to the crystalline structure can be detected as long as a sufficient number of electrons can be recorded for statistical significance. Indeed, our graphene membranes are highly stable in the electron beam at 100kV, allowing long data collection times on one region. For example, all images in Figs. 2-4 are recorded from graphene membranes after between one and three hours of irradiation (at ca. 7A/cm$^2$). Moreover, the summation of 100 consecutive CCD frames corresponds to an exposure time of 20 minutes, and distortions in the membrane during this time are below the resolution limit. This combination of a crystalline, atomically thin membrane along with the high beam stability and the absence of an amorphous background signal on the nominally clean membrane enable this unprecedented single-light-atom sensitivity in TEM. In comparison, single-walled carbon nanotubes (SWNTs) show strong deformations under the same dose and energy of electron irradiation (see Fig. 5 of Ref. [24]), probably because the cylindrical geometry allows beam-induced defects to relax via local deformations more easily.

The observation of stable and well-localized hydrogen adatoms on graphene, in spite of the irradiation and room temperature conditions, imply that these are chemisorbed rather than physisorbed atoms. Strong bonding of hydrogen to graphite is possible if the nearest carbon atom changes its bonds from sp$^2$ to sp$^3$ configuration [25 - 27], with the carbon atom displaced from the plane by about 0.36Å (Fig. 2f). Moreover, it was found [25] that hydrogen cannot bind to graphene if the carbon is confined to a plane (e.g. by strong bonding to a substrate), while an isolated membrane can deform easily to accommodate different types of bonds [28, 29]. From the observed density of hydrogen adatoms, we conclude that only about 0.3% of the carbon atoms in our graphene membrane are in an configuration with a hydrogen adatom.

Our real-time observation of molecular dynamics has important implications for chemical diffusion and reaction dynamics studies. As demonstrated above, a variety of molecular scale adsorbates become trapped on the membrane, and often detach again or decompose after a few minutes. We can observe individual alkane-type molecules and we can even follow their migration. Observing this kind of molecule in the TEM has important implications because it represents an essential ingredient of organic chemistry. It therefore appears likely that other, more complex, molecules can be observed after deposition on graphene membranes. We find that the carbon chains are sufficiently stable and localized for characterization even at room temperature, and note that these adsorbates were only trapped on the membrane after a moderate density of defects had been created by irradiation.

In conclusion, we have demonstrated that graphene membranes enable a TEM visualization of ultra-low contrast objects. The imaging of individual hydrogen and carbon adatoms and carbon chains demonstrates a new level of sensitivity that is relevant for organic materials. A key strength of the TEM is its ability to image individual entities rather than averaging over an ensemble, and direct imaging promises insights ranging from the characterization of complex chemicals and nanomaterials to biological molecules. The extremely high sensitivity that a graphene membrane in the transmission electron microscope provides with respect to adsorbates has allowed us to detect even hydrogen, demonstrating the ultimate in TEM atomic sensitivity. While the study of defects, vacancies and edges of the graphene sheet itself will provide insights for potential electronic modifications of this new material, the placement of objects on graphene membranes will enable unprecedented analysis by TEM, including electron spectroscopic analysis, and the study of molecular dynamics.

## Methods summary

Graphene sheets are prepared on oxidized silicon substrates by mechanical cleavage. After identification by optical microscopy, selected sheets are transferred to Quantifoil TEM grids (Quantifoil Micro Tools GmbH, Jena, Germany) with 1.2 μm holes. The perforated TEM support film is brought into contact with the substrate and graphene sheet by evaporating a drop of solvent. The substrate is then removed by wet chemistry, while the graphene sheet remains attached to the TEM grids (details are given in the supplementary information). TEM imaging is carried out in a JEOL 2010 microscope operated at 100kV. The sample holder is at room temperature; the actual sample temperature may differ due to electron beam heating or the nearby decontaminator cold trap. A continuous sequence of images is recorded on the CCD camera. The defocus value (60 nm) and presence of vibrations is estimated from the thin amorphous coverage that intersperses the clean areas of the graphene membrane for each frame, and ca. 5% of the frames are discarded. Then, drift-compensated summations of up to 100 frames are performed (with each frame verified for imaging parameters and vibrations) to obtain an adequate signal-to-noise ratio. Orthogonal slices through the stack of images (see supplementary information) clearly establish whether a feature of interest has been static during the entire effective exposure time, or can be used to detect interesting dynamics in the data.

## References

[1]     G. Binning, H. Rohrer, Ch. Gerber, and E. Weibel. Surface studies by scanning tunneling microscopy. *Phys. Rev. Lett.* **49,** 57-61 (1982).

[2]     J. C. H. Spence. *High-Resolution Electron Microscopy.* Oxford University Press (2003).

[3]     P. R. Buseck, J. M. Cowley, and L. Eyring. *High-Resolution Transmission Electron Microscopy.* Oxford University Press (1988).

[4]     A. V. Crewe, J. Wall, and J. Langmore. Visibility of single atoms. *Science* **168,** 1338-1340 (1970).

[5]     H. Hashimoto, A. Kumao, K. Hino, K. Hino, H. Endoh, H. Yotsumoto, and A. Ono. Visualization of single atoms in molecules and crystals by dark field electron microscopy. *Journal of Electron Microscopy* **22,** 123-134 (1973).

[6]     S. Iijima. Observation of single and clusters of atoms in bright field electron microscopy. *Optik* **48,** 193-213 (1977).

[7]     P. D. Nellist and S. J. Pennycook. Direct imaging of the atomic configuration of ultradispersed catalysts. *Science* **274,** 413-415 (1996).

[8]     P. M. Voyles, D. A. Muller, J. L. Grazul, P. H. Citrin, and H.-J. L. Gossmann. Atomic-scale imaging of individual dopant atoms and clusters in highly n-type bulk silicon. *Nature* **416** 826-829 (2002).

[9]     K. v. Benthem, A. R. Lupini, M. Kim, H. S. Baik, S. Doh, J.-H. Lee, M. P. Oxley, S. D. Findlay, L. J. Allen, J. T. Luck, and S. J. Pennycook. Three-dimensional imaging of individual hafnium atoms inside a semiconductor device. *Appl. Phys. Lett.* **87,** 034104 (2005).

[10]    P. A. Doyle and P. S. Turner. Relativistic hartree-fock x-ray and electron scattering factors. *Acta Cryst. A* **24,** 390-397 (1968).

[11]    C. Kisielowski, C. J. D. Hetherington, Y. C. Wang, R. Kilaas R, M. A. O'Keefe, A. Thust. Imaging columns of the light elements carbon, nitrogen and oxygen with sub Angstrom resolution. *Ultramicroscopy* **89,** 243-263 (2001).

[12]    C. L. Jia, M. Lentzen, and K. Urban. Atomic-resolution imaging of oxygen in perovskite ceramics. *Science* **299,** 870-873 (2003).

[13]    B. W. Smith, M. Monthioux, and D. E. Luzzi. Encapsulated C in carbon nanotubes. *Nature* **396,** 323-324 (1998).

[14]    Z. Liu, M. Koshino, K. Suenaga, A. Mrzel, H. Kataura, and S. Iijima. Transmission electron microscopy imaging of individual functional groups of fullerene derivatives. *Phys. Rev. Lett.* **96,** 088304 (2006).

[15]    Z. Lui, K. Yanagi, K. Suenaga, H. Kataura, and S. Iijima. Imaging the dynamic behaviour of individual retinal chromophores confined inside carbon nanotubes. *Nature Nanotechnology* **2,** 422-425 (2007).

[16]    A. Hashimoto, K. Suenaga, A. Gloter, K. Urita, and S. Iijima. Direct evidence for atomic defects in graphene layers. *Nature* **430,** 870-873 (2004).

[17]    K. Suenaga, H. Wakabayashi, M. Koshino, Y. Sato, K. Urita, and S. Iijima. Imaging acitve topological defects in carbon nanotbues. *Nature Nanotechnology*, **2,** 358-360 (2007).

[18]    J. C. Meyer, A. K. Geim, M. I. Katsnelson, K. S. Novoselov, T. J. Booth, and S. Roth. The structure of suspended graphene sheets. *Nature*, **446,** 60-63 (2007).

[19]    K. S. Novoselov, D. Jiang, F. Schedin, T. J. Booth, V. V. Khotkevich, S. V. Morozov, and A. K. Geim. Two-dimensional atomic crystals. *Proc. Nat. Acad. Sci.*, **102,** 10451-10453 (2005).

[20]    K. S. Novoselov, A. K. Geim, S. V. Morozov, D. Jiang, M. I. Katsnelson, L. V. Grigorieva, S. V. Dubonos, and A. A. Firsov. Two-dimensional gas of massless dirac fermions in graphene.*Nature* **438,** 197-200 (2005).

[21]    Y. Zhang, J.W. Tan, H.L. Stormer, and P. Kim. Experimental observation of the quantum hall effect and berry's phase in graphene.*Nature* **438** 201-204  (2005).

[22]    B. K. Vainshtein and Z. G. Pinsker. Opredelenie Polozheniya Vodoroda V Kristallicheskoi Reshetke Parafina. Dokl. Akad. Nauk SSSR **72**, 53-56 (1950)

[23]    H. Amara, S. Latil, Ph. Lambin, and J.-C. Charlier. Scanning tunneling microscopy fingerprints of point defects in graphene: A theoretical prediction. *Phys. Rev. B.* **76,** 115423 (2007).

[24]    B. W. Smith and E. Luzzi. Electron irradiation effects in single wall carbon nanotubes. *J. Appl. Phys.* **90**, 3509-3515 (2001).

[25]    L. Jeloaica and V. Sidis. DFT investigation of the adsorption of atomic hydrogen on a cluster-model graphite surface. *Chem. Phys. Lett.* **300,** 157-162 (1999).

[26]    X. Sha and B. Jackson. First-principles study of the structural and energetic properties of H atoms on a graphite (0001) surface. *Surf. Sci.* **496,** 318-330 (2002).

[27]    L. Hornekaer, Z. Sljivancanin, W. Xu, R. Otero, E. Rauls, I. Stensgaard, E. Laegsgaard, B. Hammer and F. Besenbacher. Metastable Structures and Recombination Pathways for Atomic Hydrogen on the Graphite (0001) Surface. *Phys Rev. Lett.* **96,** 156104 (2006).

[28]    D. W. Boukhvalov, M. I. Katsnelson, and A. I. Lichtenstein. Hydrogen on graphene: Electronic structure, total energy, structural distortions and magnetism from first-principles calculations. *Phys. Rev. B* **77,** 035427 (2007).

[29]    A. Ito, H. Nakamura, and A. Takayama. Chemical reaction between single hydrogen atom and graphene. *http://www.arxiv.org/cond-mat/0703377*, 2007.

[30]    K. Nordlund, J. Keinonen, and T. Mattila. Formation of ion irradiation induced small-scale defects on graphite surfaces. *Phys. Rev. Lett.* **77,** 699-702 (1996).

Acknowledgements.  This work was supported by the Director, Office of Energy Research, Office of Basic Energy Sciences, Materials Sciences and Engineering Division, of the U.S. Department of Energy under contract No. DE-AC02-05CH11231. AZ gratefully acknowledges support from the Miller Institute of Basic Research in Science, and CG acknowledges support from an NSF Graduate Fellowship.

# Figures

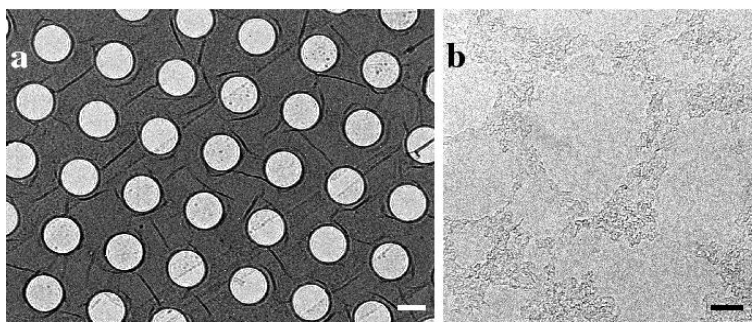

Figure 1: Graphene membrane sample as observed by TEM. (a) Low magnification overview image of a suspended graphene sheet on the perforated carbon foil. (b) High resolution close-up of a graphene membrane. We observe small, extremely clean areas with diameters of ten to fifty nanometers where no contrast is visible, separated by regions with thin amorphous adsorbates. Scale bars, 1 μm (a), 10 nm (b).

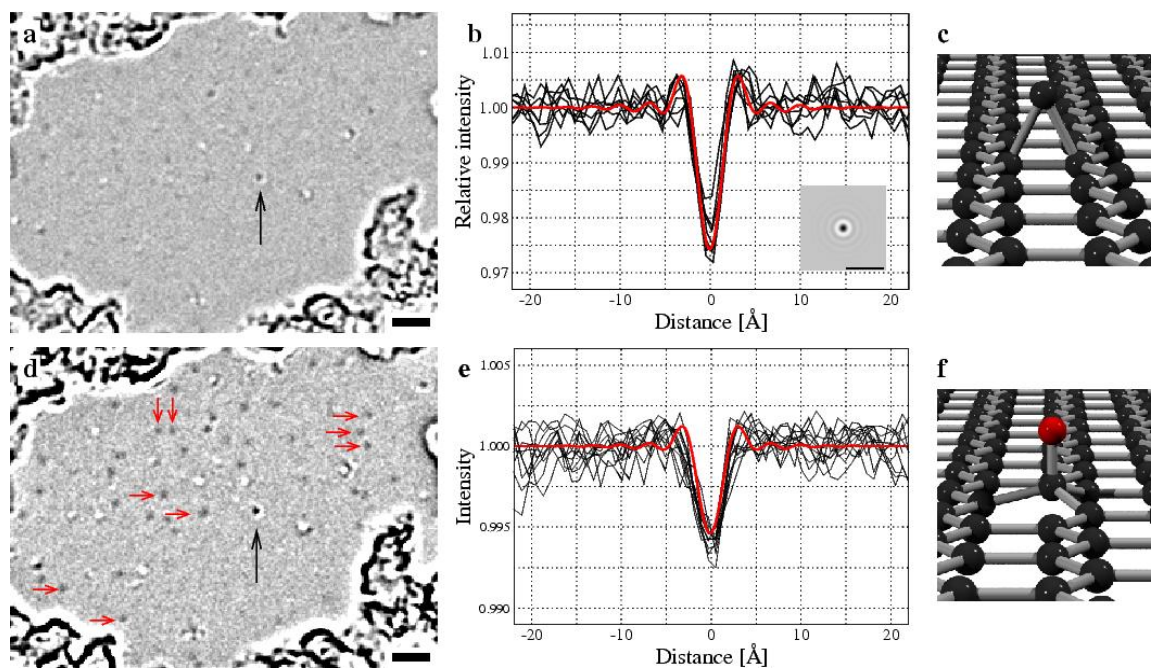

Figure 2: Adatom images. (a) Carbon adatom (black arrow). (b) Intensity profiles from several images of the carbon adatom (black), and a simulated profile (red). Inset in (b) shows the simulated image. (c) Carbon adatom configuration according to Ref. [30]. (d) Hydrogen adatoms on the same sample (dark grey spots), a selection of which are indicated by a red arrow. The profile plots are shown in (e). Black arrow in (d) is again the carbon adatom. Red line in (e) is the simulated profile for a hydrogen adatom. (f) Configuration of a chemisorbed hydrogen atom according to Ref. [25]. All scale bars are 2 nm.

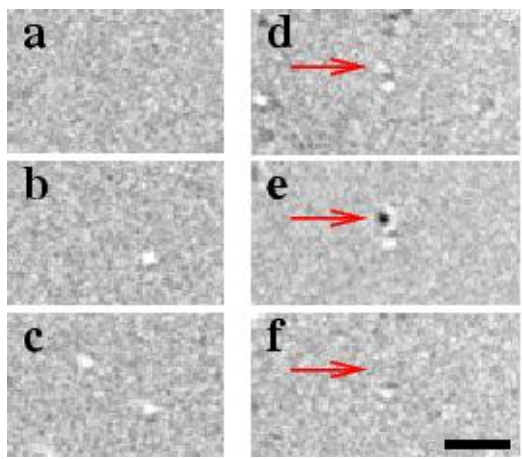

Figure 3: Dynamics of defects. (a-c) Generation of vacancies due to electron irradiation. Time between (a) and (c) is 50 minutes. (d-f) Annealing of a vacancy by interaction with an adsorbate. We observe two individual vacancies (d), and then (e) trapping of a larger adsorbate, corresponding to a mass of a few carbon atoms, on one of the defects. After ca. 5 minutes, both the adsorbate and the one vacancy disappear (f), showing that the missing carbon atom in the graphene sheet has been replaced by an atom from the adsorbate. Scale bar is 2 nm.

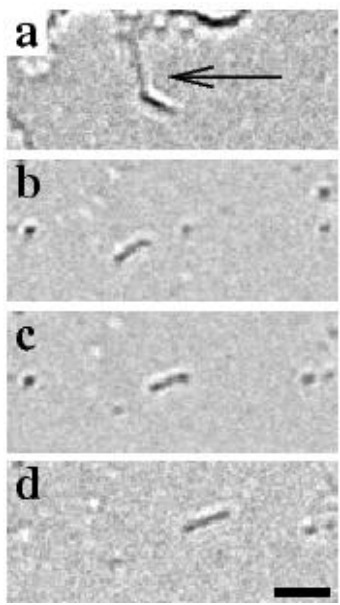

Figure 4: Molecular scale adsorbates. (a) Molecule suspended between other adsorbates (arrow). (b-d) Migration of a carbon chain, where one end remains attached in each step. This migration is also shown in the supplementary video. The contrast is in agreement with an alkane molecule. Scale bar is 2 nm.

# Imaging and Dynamics of Light Atoms and Molecules on Graphene

## Supplementary information

Jannik C. Meyer, Caglar. O. Girit, Michael F. Crommie and Alex Zettl

Department of Physics, University of California at Berkeley, and Materials Sciences Division, Lawrence Berkeley National Laboratory, Berkeley, CA 94720, U.S.A.

# 1 Supplementary Methods

## 1.1 Preparation of Graphene membranes

Our graphene membranes were prepared by cleaving graphene onto a bulk substrate using the established scotch tape method, and subsequently transferring them to a commercially available TEM grid. The actual transfer was achieved by two slightly different methods described below.

The first method to prepare graphene membranes is as follows. Graphene sheets are prepared on silicon substrates with a 300 nm silicon dioxide ($SiO_2$) layer by mechanical cleaveage and located by optical microscopy as described previously [S1 - S4]. Under the optical microscope, Quantifoil electron microscopy grids (200 Mesh Gold, 1.3 μm holes, Quantifoil Micro Tools GmbH, Jena, Germany) are placed over the graphene sheet. A drop of isopropanol is added and left to evaporate. Its surface tension during evaporation pulls the microscopy grid into close contact with the surface and the graphene sheet. The sample is then heated on a hot plate at 200°C for 5 minutes. Next, the substrate with the microscopy grid is placed in a 30% solution of semiconductor grade potassium hydroxide at room temperature. This dissolves the silicon dioxide layer while leaving the Quantifoil grid and graphene sheets intact. After a time ranging from a few minutes to a few hours, the grid along with the graphene sheets falls off from the substrate. It is then washed in water and transferred to isopropanol before drying. The graphene sheets remain suspended across the holes of the grid.

The second, slightly different, method to create graphene membranes, is as follows. Silicon substrates with a 300nm $SiO_2$ layer are coated with a few (10-30) nm of Polymethylmetacrylate (PMMA). The PMMA layer later serves as sacrificial layer while it is kept thin so as not to alter the optical properties of the substrate. Natural graphite (purchased from NGS Graphit GmbH, > 30 Mesh, source: Madagascar) is cleaved apart using sticky tape. The tape with the graphitic flakes is pressed against the coated substrates and peeled off. We found that Graphene can be obtained in this way on a variety of plastics and polymers, including PMMA, in the same way as on silicon dioxide. Again, single- and few-layer graphene sheets are located by optical microscopy. Under the optical microscope, Quantifoil electron microscopy grids (as above) are placed over the graphene sheet. A drop of isopropanol is added and left to evaporate. Its surface tension during evaporation pulls the microscopy grid into close contact with the surface and the graphene sheet. The sample is then heated on a hot plate at 200°C for 5 minutes. Next, the substrate with the microscopy grid is placed in acetone, where the PMMA layer is dissolved and the grid along with the graphene sheets is separated from the substrate. The Quantifoil grid is transferred to isopropanol and then dried, resulting in graphene sheets suspended across the holes of the grid. This second method avoids the use of acids or bases, however, the single-layer regions appear to break slightly more frequently during preparation than in the first approach.

Just before insertion into the TEM, the graphene membrane samples are again heated on a hot plate for 15 minutes at 200°C to reduce the amount of adsorbates that are present on the sample surface due to the wet preparation and due to air exposure.

## 1.2 Image simulations

Image simulations are carried out in a single slice, phase object approximation. We have used (a) the electron atomic scattering factors of Refs. [S5] and [S6, S7] (for hydrogen) in our own computer code, and (b) the independent computer code and scattering factors of Ref. [S8]. In both cases, projected atomic potentials are calculated from the scattering factors, applied to obtain the wave function at the exit face of the sample, and then convoluted with the appropriate contrast transfer function to obtain the simulated image. Although the two programs rely on a different calculation and parametrization of the atomic scattering factors, the results of both simulations are in excellent agreement. The parameters of the microscope and settings used in the simulation are Cs=1 mm, Cc=1.4 mm, Energy spread 2 eV (standard deviation), Illumination semiangle 1 mrad, Electron energy 100 keV, Objective aperture 30 mrad. Defocus in the experiment is between 60 nm and 78 nm as determined from the amorphous regions. Since the defocus uncertainty has the largest effect on the simulated contrast, we show here (in the supplementary information) a simulation each for the 60 nm and 72 nm defocus value: Within our range, 60 nm gives the minimum and 72 nm the maximum contrast. In the main article, the simulation is shown for 60 nm defocus. We have also verified that uncertainties in the other experimental parameters have negligible effects on the simulated contrast, and that misalignments of the microscope (such as astigmatisms) would produce a visible distortions in the images before affecting the contrast. Height variations across these samples (about 1 nm [S9]) do not produce any significant focus variation. Fig. S1 shows the simulated central dip in intensity for single atom images at our conditions. The range of observed values that were identified as hydrogen atoms are indicated by the dashed horizontal lines. Clearly, the values match only hydrogen and helium, with a large margin towards any heavier elements.

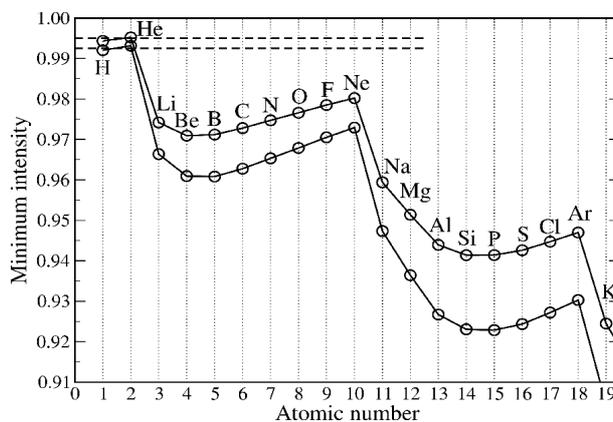

Figure S1: Simulated intensity in the central dip of single atom images, shown for the maximum and minimum contrast expected within the experimental conditions (defocus of 60 and 72 nm). Horizontal dashed lines indicate the observed contrast range of the hydrogen adatoms.

The simulations rely on scattering factors for isolated atoms. In a bonded configuration, the electron scattering factors may be different, and such an effect will be largest for light atoms. However, hydrogen in a crystalline arrangement has been detected previously in electron diffraction experiments [S10 - S15] and was found to produce at least the expected amount of scattering (and indeed more than neutral hydrogen in case of partial ionization [S11, S15]). For a C-H bond, the ratio between the scattered intensities of carbon and hydrogen was determined as 3.5 [S10], in agreement with our values.

## 1.3 Data acquisition and analysis

We record a continuous series of high-resolution TEM images of the graphene membranes using the CCD camera, typically for several hours on the same region. Individual exposures are 5s with a mean intensity value of 2500 counts per pixel (at 0.8Å pixel size). An initial drift compensation is carried out manually during acquisition by using the image shift deflectors. Drift in the beam direction is compensated by keeping the contrast transfer of the amorphous covered regions that surround the clean areas constant (judged by their Fourier transforms). After data acquisition, we verify for every frame in the sequence the defocus parameter and presence of vibrations from the power spectrum of the Fourier transform [S16, S17]. The small amorphous coverage surrounding the clean areas allow a good estimate of these parameters. Approximately 5% of the CCD frames are discarded from the sequence. The remaining set of frames is then all within 1.0 to 1.3 of the Scherzer defocus (60 nm), and free of detectable vibrations. A precise drift compensation is carried out numerically (see e.g. section 2.2 of Ref. [S18]), which again is helped by the thin amorphous coverage surrounding the clean graphene windows.

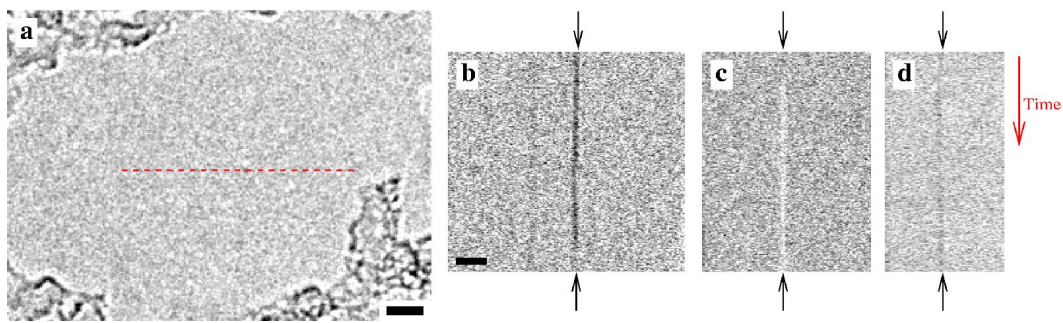

Figure S2: (a) Individual CCD frame from an image sequence. The aligned image sequence is treated as a 3D data set, and (b-c) shows orthogonal cuts for a sequence of 182 frames. For each frame, the red line in (a) (as example) corresponds to one horizontal line in panel (b). (b) Trace of a carbon adatom, clearly present throughout most of the sequence. The carbon adatom detaches close to the end of the image sequence, near the lower edge of panel (b). (c) Formation of a vacancy by knock-on damage, visible as the beginning of a white line near the upper edge. (d) Trace of a hydrogen adatom, clearly present throughout the entire image sequence. For any image averaged across several frames (a subset of the sequence) such as shown in the main article, the orthogonal cuts clearly establish which of the features have been present and static during the entire effective exposure time.

We now look at orthogonal cuts through the 3D data set that is formed by the sequence of aligned images. Fig. S2a shows an individual frame and Fig. S2b an orthogonal cut, where each horizontal line corresponds to the dashed line in Fig S2a for subsequent frames. In this example, the location of the individual carbon adatom is chosen - it is not visible in the single exposure, but the "re-sliced" data shows a continuous line showing the adatom has not moved, adsorbed or desorbed during the time of observation. In a similar way, we can detect vacancies, or indeed follow the dynamics by looking for start and end points of white and dark lines (Figs. S2b-d). The fact that all features show up as precisely vertical lines prove that the alignment precision is better than the resolution of the microscope and that no deformations in the membrane are introduced by the irradiation.

A continuous line in this data set identifies a feature that has not changed during the corresponding data acquisition time. Vacancies appear as white lines and adsorbates as dark lines. The elementary vacancy formed by knock-on damage is a single carbon atom removal. For the unrelaxed vacancy one would expect the precise opposite contrast of a carbon adatom (i.e., 2.7% contrast in a white spot). However, a relaxed vacancy (calculated for graphene in Ref. [S19]), where most nearby atoms shift towards the vacancy, results in a significantly lower signal. Using the relaxed atom positions in Ref. [S19], the simulation predicts a white spot with an intensity 2.0% above the mean value. Experimentally observed white spots that are formed during observation (therefore most likely single vacancies just after their formation) all show a white contrast of about 1.5% (Fig. S2c). It must be noted that the simulated intensity for the relaxed atom vacancy is rather sensitive to the precise atom positions, and therefore is not taken as a precise reference (the edge is more well defined, see below). It does, however, provide additional evidence that the mismatch between simulated and experimental intensities in this experiment is rather small.

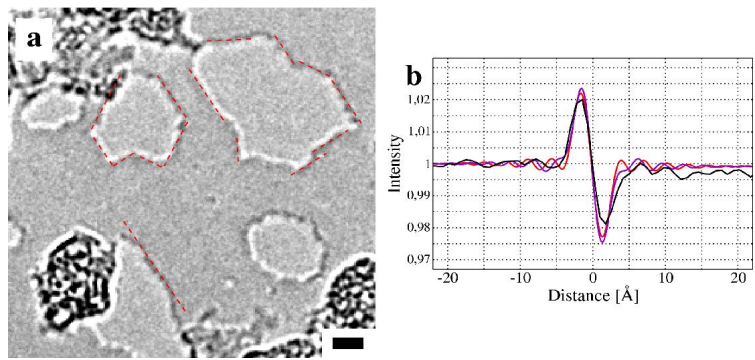

Figure S3: (a) Polygon shaped holes formed after prolonged irradiation. All red dashed lines incorporate angles that are multiples of 30°. (b) Intensity profile plot across an edge from vacuum (left) to the single-layer graphene membrane (right). The black line is a measured intensity profile, while the red and pink lines are simulations for defocus values of 60 nm and 72 nm, respectively.

A reference with known structure is provided by the edges of the membrane (Fig. S3). In particular, holes that form after prolonged irradiation show polygon shapes with all angles at multiples of 30 degrees. Therefore, we assume the edges of the holes to be armchair and zigzag lines of the graphene lattice. Independent of the orientation, the edges provide a fringe contrast that is in excellent agreement with the image simulation. In addition, a step edge between a single layer and bi-layer region of a sample was characterized and found to provide the precise same intensity fringe. The thicknesses were independently determined by electron diffraction [S9, S20].

In addition, the intensity plot across the edge of a hole confirms that the contrast here is dominated by phase contrast (elastic scattering): The dominant, symmetric fringe (+/- 2%) at the edge is evidently a phase contrast effect. It is only within the featureless region of the graphene sheet where an absorption (due to scattering outside of the aperture) can be detected: We observe about 0.3% intensity difference between empty hole and clean single layer regions, somewhat larger than predicted by the elastic scattering simulation (ca. 0.1%). Given the uncertainty in experiments and calculations for *inelastic* scattering cross sections [S21], it is not clear how much inelastic scattering will affect a single hydrogen image. However, any amount of absorption added into the simulation (by introducing a complex scattering potential) does only *increase* the contrast, and therefore can not lead to a false identification of a heavier atom as hydrogen atom.

# 2 Supplementary Discussion

## 2.1 The Stobbs factor

Experimental high-resolution lattice images of "thin" crystals are frequently reported to have a much lower contrast than simulated images. The mismatch factor in the literature ranges between 1.5 and 6 [S22 - S27]. The origin of this mismatch is still not fully understood, and most likely a combination of several effects that apply in different amounts to different types of samples. We argue here that the possible explanations that were put forward previously to explain the so-called "Stobbs factor" do not apply in our case. Before doing so, we note again that we study a one-atom thick support that is at least an order of magnitude thinner than typical "thin" TEM samples, and evidently free from any amorphous coating in the regions of interest (the regions where individual adatoms are observed). Further, we consider isolated features rather than lattice images. It must be noted that our single layer with an adatom is very well modeled in the rather simple weak phase object approximation, while almost all thicker samples require more elaborate calculations. Also, previous studies of heavier supported isolated atoms [S28] report a match to simulations better than a factor of two, in spite of a much thicker support.

First, amorphous coatings on a sample have been shown experimentally to contribute a dominant part in the contrast mismatch [S23, S24, S26]. Our region of interest is free from any such coating, with a very thin amorphous coverage only *surrounding* the clean regions. Second, electrons scattered inelastically by more than a few eV (plasmon losses) are out of focus and provide a more or less uniform background intensity that effectively reduces the contrast. This contribution scales with the sample thickness. Thus, while as much as 15% of the intensity in an unfiltered image of a 10 nm thick sample [S23] may be due to this background intensity, it will add less than 1% background intensity in our case of a single layer. Third, defects in crystalline specimen could lead to an effectively very large Debye-Waller factor in the projection of atomic colums [S22]. For a single layer / single atom observation, this does obviously not apply. Forth, vibrations of the lattice can smear out high-resolution images [S22, S27]. This is an effect that affects predominantly very-high resolution images. In our case, however, the intensity dip that shows the adatom is not significantly affected by adding any reasonable vibrations (it is already wide enough on its own). Finally, low-loss (phonon scattered) electrons have recently been shown to contribute to a lattice image that can be out of phase with the elastic image [S25], and thereby reduce the overall contrast. It was concluded that the mismatch factor due to this effect alone would be at most 1.4, and in addition, is a complicated function of sample thickness. While it is not clear how much this last effect affects a single-atom image on graphene, we note that our conclusions will not be affected even if we allow a contrast mismatch of 1.4: Fig. S1 indicates a much larger safety margin for the hydrogen identification.

Again, we point out the excellent contrast match at the edge of a graphene sheet (near a hole), such as shown in Fig. S3. These holes provide a reference that is frequently present in the same image (i.e., recorded at identical conditions) as other features such as adatoms or vacancies. Also, the width of the edge fringe is similar to that of an adatom profile, i.e., it involves a similar set of spatial frequencies.

## 2.2 Knock-on damage cross sections

The cross section for knock on damage in an electron beam is given by Seitz and Koehler [S29], here in SI units, as

$$\sigma_d = \left(\frac{Ze^2}{\epsilon_0 2 m_0 c^2}\right)^2 \frac{1-\beta^2}{4\pi\beta^4} \left\{\frac{T_m}{E_d} - 1 - \beta^2 \ln\left(\frac{T_m}{E_d}\right) + \pi \frac{Ze^2}{\hbar c}\beta \left[2\left(\frac{T_m}{E_d}\right)^{\frac{1}{2}} - \ln\left(\frac{T_m}{E_d}\right) - 2\right]\right\} \tag{1}$$

Apart from the usual constants ($e$ electron charge, $m_0$ electron mass, $c$ speed of light, $\beta=v/c$ with $v$ the velocity of the electron, $\varepsilon_0$ electric constant in vacuum, $M$ atomic mass) the expression depends on the threshold energy for displacement $E_d$, the maximum transmitted energy in a scattering event

$$T_m = \frac{2ME(E + 2mc^2)}{(M + m)^2 c^2 + 2ME}$$

(the beam energy E enters into $T_m$ and also in $\beta=v/c$), and the atomic number Z. We first determine the displacement threshold for a hydrogen atom bonded to the graphene sheet. The calculations in Refs. [S30, S31] consider two cases: A slow (adiabatic) attachment/detachment of hydrogen where the graphene lattice is allowed to relax, and a frozen carbon lattice, showing the effective potential for the hydrogen atom in case of a rapid displacement. From these results, we obtain the displacement barrier for the hydrogen atom as 0.8 eV and 1.4 eV for the two cases, respectively. Since a knock-on displacement in the electron beam is fast in comparison to lattice relaxation time scales, we use the latter value. However, the potential well for the bonded hydrogen (Fig. 3 in Ref. [S30]) is highly asymmetric: A hydrogen atom on the top surface, where knock-on displacement occurs towards the graphene sheet, will have a much higher displacement threshold than an atom at the bottom surface. We will thus assume the displacement threshold of 1.4 eV for the hydrogen at the bottom surface. For a hydrogen to traverse a graphene sheet, it was calculated in Ref. [S32] that more than 14 eV are required, and in particular, much more if the hydrogen is centered above a carbon atom. In addition, Refs. [S32, S33] calculate that the hydrogen atom will most likely chemisorb (and thus, in our case, presumably remain chemisorbed) if accelerated towards the sheet at less than about 10 eV. Although it is a somewhat simplistic approximation to use anisotropic displacement thresholds to calculate knock-on probabilities in different directions, this has previously produced reasonable results for anisotropic carbon knock-out from nanotubes [S34, S35] that were later refined in a detailed calculation [S36]. In any case, these numbers should be considered an order-of-magnitude estimate, and indicate a reasonable stability for hydrogen on the top surface.

The maximum transmitted energy for a 100kV electron displacing a hydrogen atom is $T_m=239$eV. With $E_d=1.4$eV we obtain a displacement cross section of 324 barn, and for $E_d=14$eV a value of 29 barn (1 barn = $10^{-24}$cm$^2$). This translates into an expected lifetime of 1.5 and 14 minutes, respectively, in our electron beam (ca. 7A/cm$^2$). The displacement is a statistical process, and we note again that we analyze only those adatoms that have unambiguously been in place during the entire exposure time of the averaged image. The orthogonal cuts of image sequences before averaging, as shown in Fig. S2 show clearly how long all atoms and defects have been present. Observed lifetimes of the hydrogen adatoms are on the order of 45 minutes.

It has been noted previously (from above formula) that, at very high electron energies, the displacement cross section is actually larger for higher atomic number (Z) atoms. This is because for $E_d \ll T_m$, the $Z^2$ dependence dominates over the atomic mass dependence in $T_m$. However, we note here that with relatively small displacement thresholds (on the order of 1 eV), the same is true already at our modest electron energy of 100 keV. A careful analysis of Eq. 1 shows that weakly bonded hydrogen is in fact more stable than any similar weak bonded heavier atom, such as a carbon adatoms or even inert metal atoms on a carbon film.

# 3 Supplementary video legends

## 3.1 Supplementary video #1

Dynamics of a linear molecule on a graphene membrane as in Figs. 4b-d of the main article. Horizontal field of view in the video is 10 nm.

### 3.2 Supplementary video #2

Dynamics of a carbon chain attached between larger adsorbates. Horizontal field of view is 14 nm.

# References

[S1] K. S. Novoselov, A. K. Geim, S. V. Morozov, D. Jiang, Y. Zhang, S.V. Dubonos, I.V. Grigorieva, and A.A. Firsov. Electric field effect in atomically thin carbon films. *Science*, 306:666, 2004.

[S2] K. S. Novoselov, D. Jiang, F. Schedin, T. J. Booth, V. V. Khotkevich, S. V. Morozov, and A. K. Geim. Two-dimensional atomic crystals. *Proc. Nat. Acad. Sci.*, 102:10451, 2005.

[S3] Y. Zhang, J.W. Tan, H.L. Stormer, and P. Kim. Experimental observation of the quantum hall effect and berry's phase in graphene. *Nature*, 438:201, 2005.

[S4] K. S. Novoselov, A. K. Geim, S. V. Morozov, D. Jiang, M. I. Katsnelson, L. V. Grigorieva, S. V. Dubonos, and A. A. Firsov. Two-dimensional gas of massless dirac fermions in graphene. *Nature*, 438:197, 2005.

[S5] P. A. Doyle and P. S. Turner. Relativistic hartree-fock x-ray and electron scattering factors. *Acta Cryst. A*, 24:390, 1968.

[S6] L.-M. Peng, G. Ren, S. L. Dudarev, and M. J. Whelan. Robust parameterization of elastic and absorptive electron atomic scattering factors. *Acta Cryst. A*, 52:257, 1996.

[S7] L.-M. Peng. Electron atomic scattering factors and scattering potentials of crystals. *micron*, 30:625, 1999.

[S8] E. J. Kirkland. *Advanced Computing in Electron Microscopy*. Plenum Press, New York, 1998.

[S9] J. C. Meyer, A. K. Geim, M. I. Katsnelson, K. S. Novoselov, T. J. Booth, and S. Roth. The structure of suspended graphene sheets. *Nature*, 446:60, 2007.

[S10] B. K. Vainshtein and Z. G. Pinsker. Opredelenie polozheniya vodoroda v kristallicheskoi reshetke parafina. *Dokl. Akad. Nauk. SSSR*, 72:53, 1950.

[S11] J. M. Cowley. Electron diffraction study of the hydrogen bonds in boric acid. *Nature*, 171:440, 1953.

[S12] K. Shimaoka. Determination of hydrogen position in cubic ice by electron diffraction. *Acta Cryst.*, 10:710, 1957.

[S13] B. K. Vainshtein. *Modern Crystallography*. Springer, New York, 1964.

[S14] B. K. Vainshtein, B. Zvyagin, and A. Avilov. *Electron diffraction techniques*. Oxford University Press, New York, 1992.

[S15] J. M. Cowley. *Diffraction Physics*. Elsevier, Amsterdam, The Netherlands, 1995.

[S16] J. C. H. Spence. *High-Resolution Electron Microscopy*. Oxford University Press, 2003.

[S17] P. R. Buseck, J. M. Cowley, and L. Eyring. *High-Resolution Transmission Electron Microscopy*. Oxford University Press, 1988.

[S18] C. T. Koch. A flux-preserving non-linear inline holography reconstruction algorithm for partially coherent electrons. *Ultramicroscopy*, 108:141, 2008.

[S19] H. Amara, S. Latil, Ph. Lambin, and J.-C. Charlier. Scanning tunneling microscopy fingerprints of point defects in graphene: A theoretical prediction. *Phys. Rev. B.*, 76:115423, 2007.

[S20] J. C. Meyer, A. K. Geim, M. I. Katsnelson, K. S. Novoselov, D. Obergfell, S. Roth, C. Girit, and A. Zettl. On the roughness of single- and bi-layer graphene membranes. *Solid State Communications*, 143:101, 2007.

[S21] R. F. Egerton. *Physical Principles of Electron Microscopy.* Springer, New York, 2005.

[S22] M. J. Hytch and W. M. Stobbs. Quantitative comparison of high resolution tem images with image simulations. *Ultramicroscopy*, 53:191, 1994.

[S23] C. B. Boothroyd. Why don't high-resolution simulations and images match? *J. of Micr.*, 190:99, 1998.

[S24] C. B. Boothroyd. Quantification of lattice images: the contribution from diffuse scattering. *J. El. Micr.*, 51:S279, 2002.

[S25] C. B. Boothroyd and R. E. Dunin-Borkowski. The contribution of phonon scattering to high-resolution images measured by off-axis electron holography. *Ultramicroscopy*, 98:115, 2004.

[S26] A. Howie. Hunting the stobbs factor. *Ultramicroscopy*, 98:73, 2004.

[S27] K. Du, K. v. Hochmeister, and F. Phillipp. Quantitative comparison of image contrast and pattern between experimental and simulated high-resolution transmission electron micrographs. *Ultramicroscopy*, 107:281, 2007.

[S28] S. Iijima. Observation of single and clusters of atoms in bright field electron microscopy. *Optik*, 48:193, 1977.

[S29] F. Seitz and J. Koehler. *Solid State Physics*, volume 2. Academic, New York, 1956.

[S30] X. Sha and B. Jackson. First-principles study of the structural and energetic properties of H atoms on a graphite (0001) surface. *Surf. Sci.*, 496:318, 2002.

[S31] X. Sha, B. Jackson, and D. Lemoine. Quantum studies of eley-rideal reactions between H atoms on a graphite surface. *J. Chem. Phys.*, 116:7158, 2002.

[S32] A. Ito, A. Takayama, and H. Nakamura. Hydrogen adsorption of back side of graphene. *http://www.arxiv.org/cond-mat/0710.2878*, 2007.

[S33] A. Ito, H. Nakamura, and A. Takayama. Chemical reaction between single hydrogen atom and graphene. *http://www.arxiv.org/cond-mat/0703377*, 2007.

[S34] V. H. Crespi, N. G. Chopra, M. L. Cohen, A. Zettl, and S. G. Louie. Anisotropic electron-beam damage and the collapse of carbon nanotubes. *Phys. Rev. B*, 54:5927, 1996.

[S35] B. W. Smith and E. Luzzi. Electron irradiation effects in single wall carbon nanotubes. *J. Appl. Phys.*, 90:3509, 2001.

[S36] A. Zobelli, A. Gloter, C. P. Ewels, G. Seifert, and C. Colliex. Electron knock-on cross section of carbon and boron nitride nanotubes. *Phys. Rev. B*, 75:245402, 2007.